# Modeling National Trends on Health in the Philippines Using ARIMA

Florence Jean B Talirongan*, Hidear Talirongan and Markdy Y Orong

*Misamis University, Ozamiz City, Philippines*

## Abstract

Health is a very important prerequisite in people's well-being and happiness. Several studies were more focused on presenting the occurrence on specific disease like forecasting the number of dengue and malaria cases. This paper utilized the time series data for trend analysis and data forecasting using ARIMA model to visualize the trends of health data on the ten leading causes of deaths, leading cause of morbidity and leading cause of infants' deaths particularly in the Philippines presented in a tabular data. Figures for each disease trend are presented individually with the use of the GRETL software. Forecasting results of the leading causes of death showed that Diseases of the heart, vascular system, accidents, Chronic lower respiratory diseases and Chronic Tuberculosis (all forms) showed a slight changed of the forecasted data, Malignant neoplasms showed unstable behavior of the forecasted data, and Pneumonia, diabetes mellitus, Nephritis, nephrotic syndrome and nephrosis and certain conditions originating in perinatal showed a decreasing patterns based on the forecasted data. In terms of the predicted health disease on leading causes of morbidity, it is evident that acute respiratory infection showed a slight decrease pattern, bronchitis and TB other forms and TB respiratory showed an increase patterns, influenza showed unstable pattern, acute watery diarrhea and acute febrile illness showed a slightly decreasing patterns. On the other hand, in terms of the predicted results of health disease on leading causes of infant's deaths, it is evident that all causes of death, pneumonia of newborn, Neonatal aspiration syndromes and other congenital malformations showed a slight decrease of prediction patter. Further, Bacterial sepsis of newborn, the respiratory distress of newborn, Disorder related to short gestation and low birth weight, not elsewhere classified, congenital pneumonia and Diarrhea and gastroenteritis of presumed infectious origin showed an increasing pattern of prediction. Future research work will focus on the measles and chickenpox trends and predictions utilizing another algorithm for trend analysis and forecasting.

**Keywords:** ARIMA • Health • Diseases • Forecasting • Trend analysis

## Introduction

People weigh high value on health since it is a core element in people's well-being and happiness [1] to the extent of setting it as a priority in the governmental and societal agenda [2]. Health is considered as an important prerequisite in reaching person's goals and aspirations, and contributory factor to development of societal undertakings [1,3,4].

The World Health Organization (WHO) battled underfunded diseases like AIDS/HIV and tuberculosis and issued International Health Regulations to make medical standards consistent and stop epidemics in their tracks [5]. In the United States, leading causes of infant, neonatal and postneonatal death in rank order which include diseases of heart; malignant neoplasms; chronic lower respiratory diseases; cerebrovascular diseases; accidents (unintentional injuries); alzheimer's disease; diabetes mellitus; influenza and pneumonia; nephritis, nephrotic syndrome and nephrosis; and intentional self-harm (suicide) [6]. In Asia Pacific regions like Thailand and China account for a significant burden of global poverty and disease. Indeed, approximately one-third of the world's intestinal helminthiases, most of the food-borne trematode infections, one-half of the active trachoma infections and a significant number of cases of lymphatic filariasis (LF), schistosomiasis and arboviral infections occur in the region [7]. Urban healths in developing countries were taken care of by WHO in terms of global overview on health and the cities, policies and health status and health environment [4,5,8]. However, it was an extraordinary opportunity to create a sustained global movement against premature death and preventable morbidity and disability from NCDs, mainly heart disease, stroke, cancer, diabetes, and chronic respiratory disease. The increasing global crisis in NCDs is a barrier to development goals including health [8,9].

Reflecting the information that is most readily available in the Philippines have placed on diseases like diabetes, tropical diseases, gastroduodenal diseases, dengue, lung disease, foot and mouth disease and allergenic rhinitis [10-13].

It was observed that several studies were more focused on the trend of individual disease however no further studies concentrated on the trends of health data. The objective of this research is to visualize the national trends of health data particularly in the Philippines. This motivates the researcher to identify the most and least vulnerable disease in the country. The trend analysis utilized time series and data forecasting using autoregressive integrated moving average (ARIMA).

### Theoretical framework

**Review of related literature:** Several literature and studies support the use of ARIMA model as a forecasting tool in predicting diseases. Li et al., [14] applied ARIMA in the incidence of hemorrhagic fever with renal Syndrome (HFRS) in China during 1986 to 2009 which fit to the given study and can be applied to future forecasting on prevention and control. Yu et al., [15] constructed ARIMA model to forecast the number of HIV infections from 2013 to 2017 in Korea. Research done by Midekisa et al., [16] and Zhang et al., [17] in Ethiopia and China used ARIMA to quantify the relationship between malaria cases. Another study of Naiman et al., [18] used ARIMA model to test the relation between smoking bans and admission rates that lead to cardiovascular and respiratory conditions. Baker-Austin et al., [19] illustrated associations between environmental changes to the emergence of vibrio infections and forecasted the risk of infections. Time series analysis of dengue incidence was studied by Gharbi et al., [20] in Guadeloupe, French West Indies. ARIMA model was used to predict the occurrence of dengue epidemics which was also studied Wongkoon [21] in Thailand and Johansson et al., [22] in Mexico. The impact of climate change on dengue transmission in the Asia-Pacific region was also examined by Banu [23]. Soebiyanto [24] analyzed the role of climatic variables as input series on influenza transmission in two regions: Hong Kong (China) and Maricopa County (Arizona, USA) where the influenza cases depend on its past values and error signal. Razvodovsky [25] found out that alcohol is an important contributor to the liver cirrhosis mortality rate in Russia.

*Address for Correspondence:* Florence Jean B Talirongan, Misamis University, Ozamiz City, Philippines, Tel: +09054137874, E-mail: badilles.fj@gmail.com

## Materials and Methods

### Materials

The data used in the study are the statistical data taken from the Philippines in Figures of the Philippine Statistics Authority (PSA) from the year 2012 up to 2016. There were several areas presented in the statistics but this study will highlight the Health comprising three areas of data on ten leading causes of death, leading causes of morbidity and leading causes of infant's deaths. These data were taken from the sources like PSA, Family Planning Survey, National Demographic and Health Survey, Department of Health and Department of Social Welfare and Development. The data were taken from the books of Philippines in Figures 2017 to 2018. However, the book only reflected data from 2012 up to 2016 which will be used for trend analysis and data forecasting on the three areas.

### Methods

The study utilized Autoregressive Integrated Moving Average (ARIMA) model employed in many fields to construct models for forecasting time series [14,26]. ARIMA (p,d,q) algorithm is used to forecast the data pattern of diseases for the next fourteen years. Time series predictions are based on changes over time in historical data sets and can produce mathematical models by using statistical data that can be extrapolated [27,28]. The ARIMA (p,d,q) model is defined as follows:

$$X_t = \Phi_1 X_{t-1} + \ldots + \Phi_p X_{t-p} + a_t - \Theta_1 a_{t-1} - \ldots - \Theta_q a_{t-q} \quad (1)$$

Where, $\Phi$'s (phis) represents the autoregressive parameters to be estimated, $\Theta$'s (thetas) are the moving average parameters to be determined, the original series is represented by X's, and the a's are the unknown random errors which are assumed to follow the normal probability distribution.

Three steps were performed to predict the incidence of leading causes of death, morbidity and infants by using the ARIMA-related modules. Model identification used autocorrelation analysis and partial autocorrelation analysis methods to analyze any random, stationary, and seasonal effects on the time series data. The researcher prepared a stationary time series by considering the differences and then determined plausible models on the basis of an autocorrelogram and a partial autocorrelogram. Lastly, the parameter estimation and model testing were used to compare the plausible models obtained, and we selected the most appropriate model. Finally, we conducted predictive analysis. The study used GRETL (Gnu Regression, Econometrics and Time-series Library) software for plotting the graphs and analysis of the data sets. Figure 1 presents the architectural design in predicting incidences of leading causes of death, morbidity and infant's death.

## Results and Discussion

In trend analysis of the three areas of data on ten leading causes of death, leading causes of morbidity and leading causes of infant's deaths, the GRETL software was utilized. Tables 1-3 present the raw data on causes of death covering year 2012 until 2016.

### Forecasting

ARIMA model was used to forecast the occurrences of each health disease. The study will forecast for the next fourteen years using the available historical data. Figures 2-4 show the graph of the predicted health disease on ten leading causes of death from 2017 to 2030 having 95% interval.

It is evident that the predicted cause of death which is the diseases of the heart from 2017 to 2030 showed a very slight change forecasted data pattern. Result signifies that heart diseases cases in the country are not stable in terms of its occurrences based on the forecasted data. This result is still true with disease with the vascular system, accidents, chronic lower respiratory diseases and Chronic Tuberculosis (all forms). Hence, a continuous mitigation strategy by the Department of Health should be conducted to mitigate the occurrence of the identified disease in the country.

It is evident that the predicted cause of death which is malignant neoplasms from 2017 to 2030 showed unstable behavior of data pattern. Result signifies that there will be years in the future that the number of occurrence of the disease will increase and it will eventually decrease. Hence, regular monitoring of its occurrence and regular conduct of mitigation plans should be observed by the concerned agency in the country.

The predicted cause of death which is the pneumonia from 2017 to 2030 showed a slight decreasing pattern based from the historical data. The result showed similarity with diabetes mellitus, Nephritis, nephrotic syndrome and nephrosis and certain conditions originating in perinatal. A result signifies that the existing intervention plans of the Department of Health in mitigating the occurrence of the diseases are effective. A continuous support to the health agency from the government should be in place at all times to continually lessen the occurrences.

Based on the identified causes of death, it is evident that the heart diseases were the leading among others and it was found out that certain conditions originating in perinatal period as one of the causes of death was found as the least cause of death based on the data. Results were supported in the reported causes of most death in the Philippines showed in the health data organization website (healthdata.org, 2019) [29].

Figures 4-8 show the graph of the predicted health disease on leading causes of morbidity from 2017 to 2030.

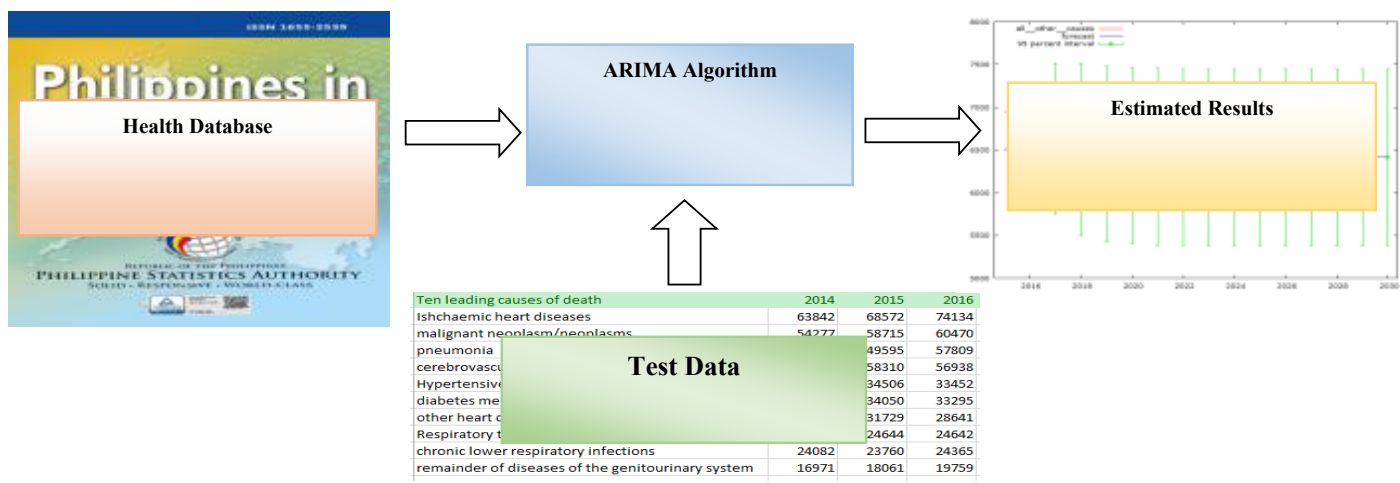

**Figure 1.** Predicting diseases trends.

Table 1. Raw data on ten leading causes of death.

| Ten Leading Causes of Death | Year | | | | |
|---|---|---|---|---|---|
| | 2012 | 2013 | 2014 | 2015 | 2016 |
| Diseases of the heart | 112581 | 118740 | 125906 | 68572 | 74134 |
| Diseases of the vascular system | 68826 | 68325 | 69913 | 58715 | 60470 |
| malignant neoplasms | 50507 | 53601 | 56219 | 49595 | 57809 |
| pneumonia | 50144 | 53101 | 54877 | 58310 | 56938 |
| accidents | 36375 | 40071 | 43853 | 34506 | 33452 |
| diabetes mellitus | 22910 | 27064 | 31687 | 34050 | 33295 |
| chronic lower respiratory diseases | 24275 | 23867 | 52114 | 31729 | 28641 |
| tuberculosis, all forms | 22693 | 23216 | 24929 | 24644 | 24642 |
| nephritis, nephrotic syndrome and nephrosis | 13555 | 14954 | 15359 | 23760 | 24365 |
| certain conditions originating in perinatal period | 11374 | 10436 | 10174 | 18061 | 19759 |

Table 2. Raw data on leading causes of morbidity.

| Leading Causes of Morbidity | Year | | | | |
|---|---|---|---|---|---|
| | 2012 | 2013 | 2014 | 2015 | 2016 |
| Acute respiratory infection | 2793066 | 2174740 | 1445320 | 2115018 | 3080343 |
| ALTRI and pneumonia | 569122 | 674597 | 488415 | 474406 | 786085 |
| Hypertension | 512604 | 410432 | 475693 | 601173 | 886203 |
| Bronchitis | 338789 | 249173 | 204086 | 202343 | 200176 |
| Influenza | 232584 | 149777 | 172683 | 147400 | 216074 |
| Urinary tract infection | 276442 | 235446 | 213666 | 298200 | 288588 |
| Acute watery diarrhea | 235110 | 74876 | 91202 | 130246 | 139700 |
| TB respiratory | 93094 | 70053 | 32335 | 62396 | 87422 |
| Acute febrile illness | 85471 | | | 55759 | |
| Dengue fever | 44172 | 53750 | 26077 | 69532 | 56487 |
| TB other forms | | 30971 | 25727 | | |

Table 3. Raw data on leading causes of infant's deaths.

| Leading Causes of Infant's Death | Year | | | | |
|---|---|---|---|---|---|
| | 2012 | 2013 | 2014 | 2015 | 2016 |
| All causes | 22283 | 22254 | 21992 | 20750 | 21874 |
| Bacterial sepsis of new-born | 3669 | 3156 | 2731 | 2157 | 2136 |
| Pneumonia | 2792 | 2738 | 3146 | 2370 | 2885 |
| Respiratory distress of new-born | 2414 | 2497 | 2347 | 2276 | 2263 |
| Congenital malformation of the heart | 1452 | 1356 | 1383 | 1398 | 1407 |
| Disorder related to short gestation and low birth weight, not elsewhere classified | 1455 | 1422 | 1466 | 1278 | 1202 |
| Congenital pneumonia | 1115 | 989 | 728 | 625 | 734 |
| Neonatal aspiration syndromes | 1104 | 994 | 969 | 1036 | 1196 |
| Intrauterine hypoxia and birth asphyxia | 906 | 871 | 838 | 802 | 832 |
| Other congenital malformations | 883 | 896 | 895 | 1030 | 1000 |
| Diarrhea and gastroenteritis of presumed infectious origin | 911 | 843 | 901 | 824 | 1328 |
| All other causes | 5582 | 6492 | 6588 | 6954 | 6891 |

The predicted cause of morbidity which is the acute respiratory infection from 2017 to 2030 showed a lightly decreasing pattern. The prediction result signifies that the forecasted data still on range based on the historical data. The result is similar with ALTRI and pneumonia and Hypertension.

The predicted cause of morbidity which is the bronchitis from 2017 to 2030 showed an increasing pattern. Further, other results showed similarity with bronchitis such as the TB other forms and TB respiratory. The prediction results signify that there was an on-going increase of the occurrences of bronchitis in the Philippines.

The predicted cause of morbidity which is the influenza from 2017 to 2030 showed unstable pattern of the case from the year 2016 to 2020. On the other hand, there is a stable prediction pattern of the case from the year 2023 to 2030.

The predicted cause of morbidity which is the urinary tract infection from 2017 to 2030 showed a stable number of forecasted cases based on the pattern presented on the graph. Further, the result is similar with the dengue fever. The prediction signifies that the forecasted data reach the maximum number of occurrences in the Philippines.

The predicted cause of morbidity which is the acute watery diarrhea from 2017 to 2030 showed a lightly decreasing pattern similar with the acute febrile illness. The prediction signifies that the forecasted data reach the maximum number of occurrences in the Philippines. Figures 9-11 show the graph of the predicted health disease on leading causes of infant's deaths from 2017 to 2030.

The predicted cause of infant's deaths which is the all causes from 2017 to 2030 showed a lightly decreased pattern which is similar with the results of Pneumonia of newborn, Neonatal aspiration syndromes, other congenital malformations and all other causes of death. The prediction results of the mentioned causes signify that the forecasted data reach the maximum number of occurrences in the Philippines.

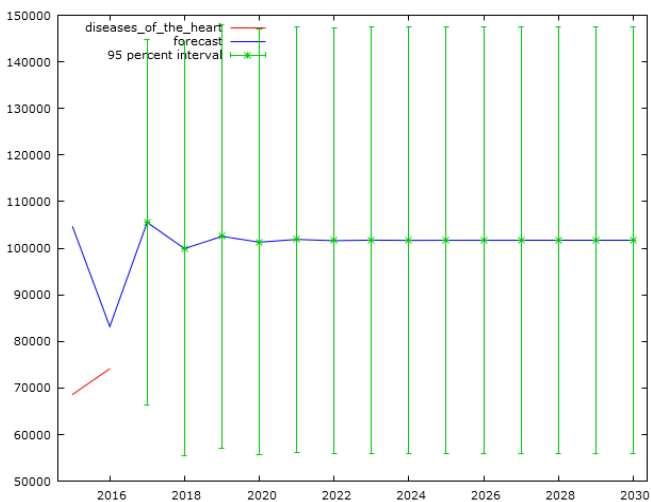

**Figure 2.** Forecasted ten leading causes of death - Diseases of the heart from 2017 to 2030.

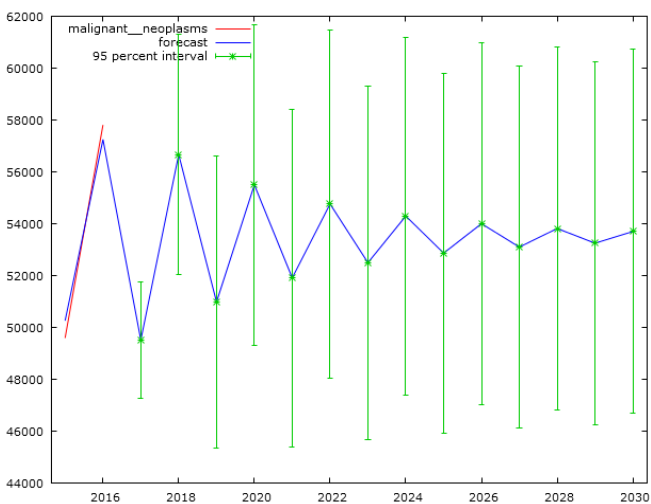

**Figure 3.** Forecasted ten leading causes of death - Malignant neoplasms from 2017 to 2030.

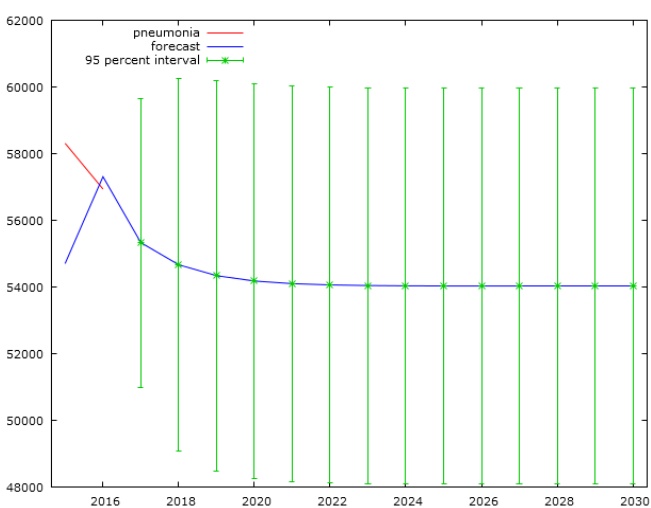

**Figure 4.** Forecasted ten leading causes of death - Pneumonia from 2017 to 2030.

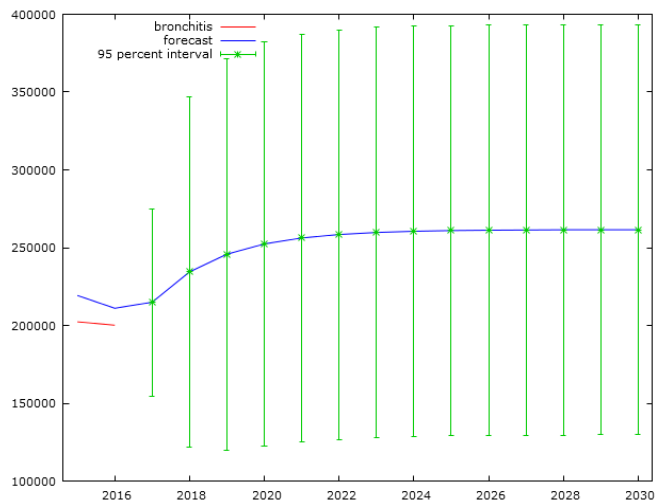

**Figure 5.** Forecasted leading causes of morbidity - Bronchitis from 2017 to 2030.

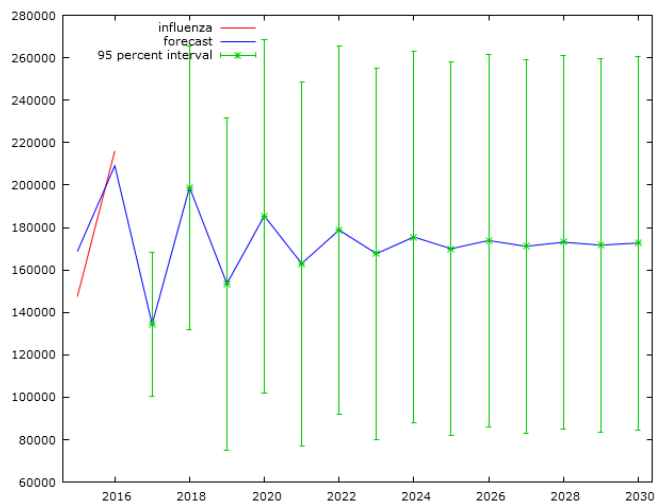

**Figure 6.** Forecasted leading causes of morbidity - Influenza from 2017 to 2030.

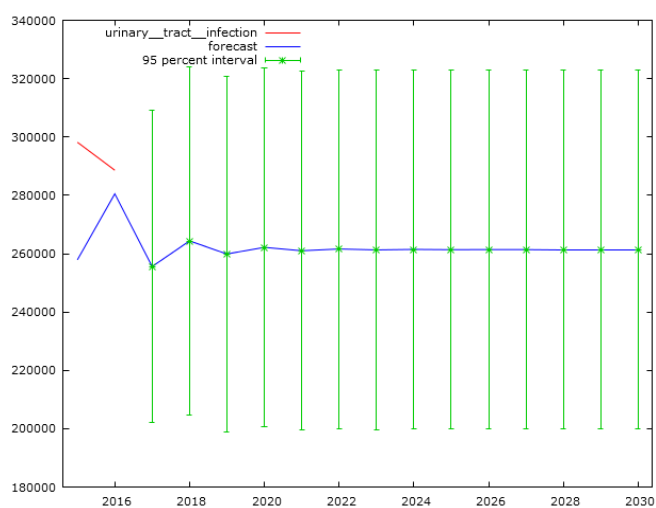

**Figure 7.** Forecasted leading causes of morbidity – Urinary tract infection from 2017 to 2030.

The predicted cause of infant's deaths which is the bacterial sepsis from 2017 to 2030 showed a gradual increase of the forecasted data based on the pattern showed in the graph similar with the results of the respiratory distress of newborn, Disorder related to short gestation and low birth weight, not elsewhere classified, congenital pneumonia and Diarrhea and gastroenteritis of presumed infectious origin. Hence, intervention plans are needed to be in place by the Department of Health to mitigate the future occurrences of the identified diseases.

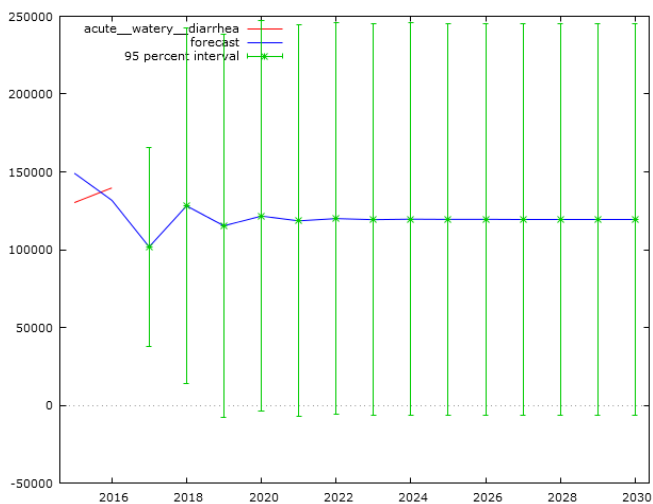

**Figure 8.** Forecasted leading causes of morbidity - Acute watery diarrhea from 2017 to 2030.

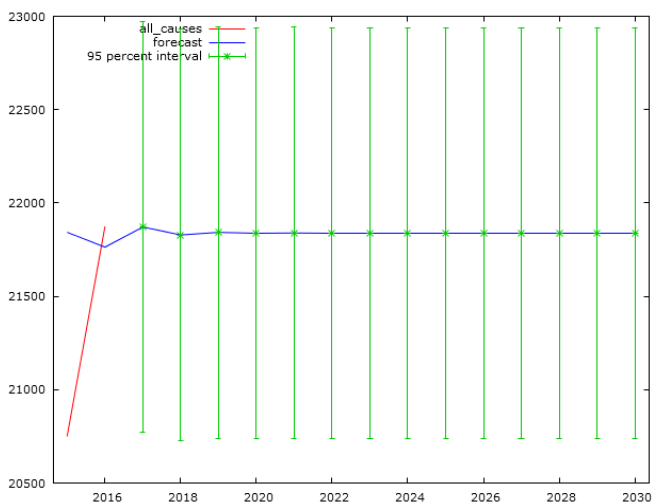

**Figure 9.** Forecasted leading causes of infant's deaths - all causes from 2017 to 2030.

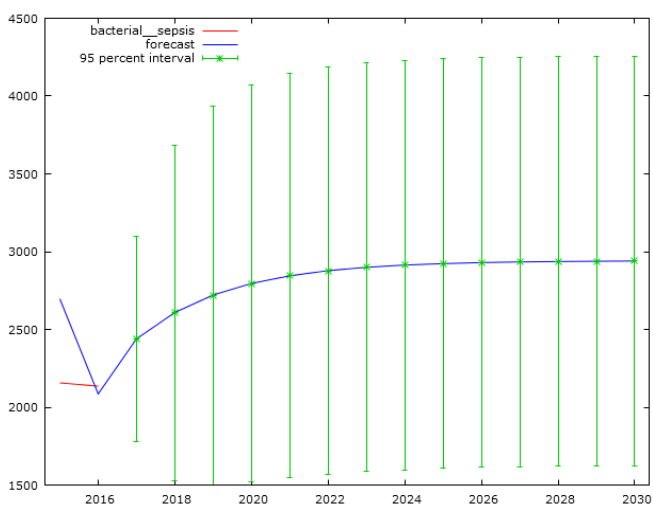

**Figure 10.** Forecasted leading causes of infant's deaths- Bacterial sepsis of newborn from 2017 to 2030.

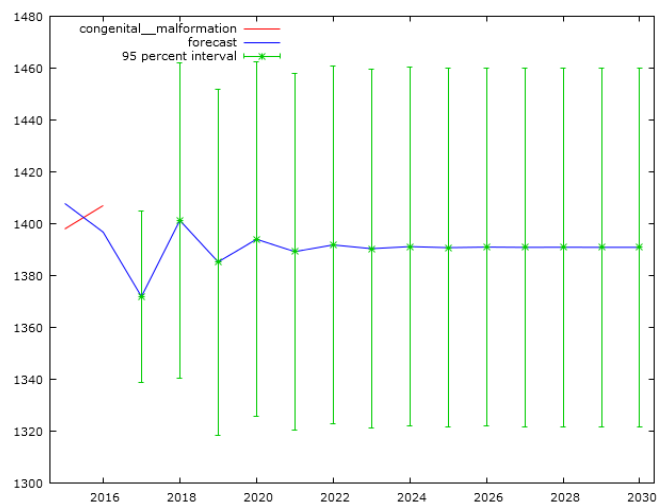

**Figure 11.** Forecasted leading causes of infant's deaths - Congenital malformation of the heart from 2017 to 2030.

The predicted cause of infant's deaths which is the congenital malformation of the heart from 2017 to 2020 showed an unstable pattern similar with the result of Intrauterine hypoxia and birth asphyxia. Moreover, it is evident that there was a stable behavior of data based on the forecast from 2021 to 2030.

## Conclusion and Recommendations

Having trend analysis and forecasting on each disease per area is very useful in foreseeing future incidences. Moreover, predicting its occurrences can give insights to the health administration in designing preventive measures against the possible spread of diseases. In the study, ARIMA (1,0,1) model helps in predicting the increasing and decreasing pattern of diseases in the community.

In the ten leading causes of death, heart disease was found as leading among others in terms of its forecasted data. On the other hand, in terms of morbidity, the acute respiratory infection was found as leading disease. Further, in the cause of disease of infant it is evident that all identified causes were found as the possible reasons of infant's death.

Future research work will focus on the measles and chickenpox trends and predictions utilizing another algorithm for trend analysis and forecasting.